# Implementing and evaluating far-field 3D X-ray diffraction at the I12 JEEP beamline, Diamond Light Source

James A. D. Ball, Anna Kareer, Oxana V. Magdysyuk, Stefan Michalik, Anastasia Vrettou, Neal Parkes, Thomas Connolley and David M. Collins

# Implementing and evaluating far-field 3D X-ray diffraction at the I12 JEEP beamline, Diamond Light Source

James A. D. Ball,[a,b] Anna Kareer,[c] Oxana V. Magdysyuk,[b] Stefan Michalik,[b] Anastasia Vrettou,[a] Neal Parkes,[a] Thomas Connolley[b] and David M. Collins[a]\*

[a]School of Metallurgy and Materials, University of Birmingham, Edgbaston, Birmingham B15 2TT, United Kingdom, [b]Diamond Light Source Ltd, Harwell Science and Innovation Campus, Didcot OX11 0DE, United Kingdom, and [c]Department of Materials, University of Oxford, Oxford OX1 3PH, United Kingdom.
\*Correspondence e-mail: d.m.collins@bham.ac.uk

Three-dimensional X-ray diffraction (3DXRD) is shown to be feasible at the I12 Joint Engineering, Environmental and Processing (JEEP) beamline of Diamond Light Source. As a demonstration, a microstructurally simple low-carbon ferritic steel was studied in a highly textured and annealed state. A processing pipeline suited to this beamline was created, using software already established in the 3DXRD user community, enabling grain centre-of-mass positions, orientations and strain tensor elements to be determined. Orientations, with texture measurements independently validated from electron backscatter diffraction (EBSD) data, possessed a ∼0.1° uncertainty, comparable with other 3DXRD instruments. The spatial resolution was limited by the far-field detector pixel size; the average of the grain centre of mass position errors was determined as $\pm\sim 80$ µm. An average per-grain error of $\sim 1 \times 10^{-3}$ for the elastic strains was also measured; this could be reduced in future experiments by improving sample preparation, geometry calibration, data collection and analysis techniques. Application of 3DXRD onto I12 shows great potential, where its implementation is highly desirable due to the flexible, open architecture of the beamline. User-owned or designed sample environments can be used, thus 3DXRD could be applied to previously unexplored scientific areas.

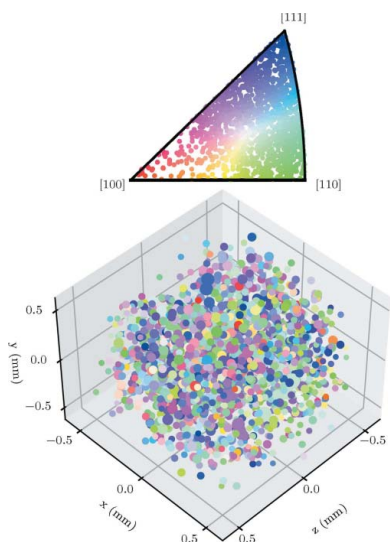

## 1. Introduction

Three-dimensional X-ray diffraction (3DXRD) is a materials characterization technique developed by the Risø National Laboratory (Poulsen et al., 1997). 3DXRD expands upon the capabilities of traditional Debye–Scherrer diffraction experiments by providing per-grain orientation, position and strain state information for a large number of simultaneously diffracting grains. Typically, Debye–Scherrer diffraction in transmission mode through a polycrystalline sample yields diffraction rings at specific values of $2\theta$ on a 2D detector. These rings can be used to determine atomic planar spacing (via Bragg's law), average powder strain (from peak shifts) and texture (variation of cone brightness with azimuthal angle on detector). If a small number (<1000) of grains are illuminated at one time, the Debye–Scherrer rings become individual spots corresponding to each grain. In 3DXRD, the sample is rotated about the vertical axis with multiple diffraction detector images taken, yielding different spot patterns on the detector. With reconstruction algorithms, per-grain position, orientation, phase and elastic strain data can be extracted for a large number of simultaneously illuminated grains at a time (Poulsen, 2012), provided that spot overlap on the detector is avoided.

3DXRD has since been implemented at several synchrotrons, including the Advanced Photon Source (Lienert *et al.*, 2011; Shade *et al.*, 2015), Cornell High Energy Synchrotron Source (Nygren *et al.*, 2020), SPring-8 (Hayashi *et al.*, 2015), PETRA III (Hegedüs *et al.*, 2019) and ESRF (Jensen *et al.*, 2006; Poulsen, 2012). 3DXRD is currently being utilized to produce significant developments in the field of polycrystalline deformation characterization, including explorations into grain neighbourhood effects during applied strain (Abdolvand *et al.*, 2018), *in situ* twinning and de-twinning during cyclic fatigue in magnesium (Murphy-Leonard *et al.*, 2019), and grain recrystallization and growth around a tin whisker (Hektor *et al.*, 2019). These advances are possible thanks to the ability of 3DXRD to quickly determine per-grain position, orientation and lattice strain information for a large number of simultaneously illuminated grains in millimetre-scale samples.

The I12 beamline at the Diamond Light Source operates at 53–150 keV which presently offers several X-ray techniques including monochromatic diffraction, small-angle X-ray scattering, energy-dispersive diffraction, radiography and tomography (Drakopoulos *et al.*, 2015). The facility has an open architecture, allowing users to bring their own sample environments with flexible size and complexity. The beamline aims for the broadest possible compatibility with different types of *in situ* processing equipment (Drakopoulos *et al.*, 2015), facilitating a wide range of novel experiments such as *in situ* diffraction with custom biaxial deformation machines (Collins *et al.*, 2015), evaluating internal strain on a connecting rod in a running motorcycle engine (Baimpas *et al.*, 2013), and measuring dynamic contact strain in a rotating ball-bearing (Mostafavi *et al.*, 2017). Utilizing this operational flexibility with 3DXRD, with the possibility of combining this with other techniques on I12, offers a tantalizing opportunity for the investigation of previously unexplored science.

To determine the viability of successfully performing a 3DXRD study at the I12 beamline, a 3DXRD experiment was performed on DX54 steel, a single-phase ferritic steel. This microstructurally simple alloy was used to develop the 3DXRD data acquisition and processing pipeline at Diamond Light Source – a necessary step before attempting the study of more complex systems or *in situ* studies of dynamic behaviour. As the DX54 alloy has been studied on the I12 beamline in previous studies along with other complimentary characterization (Collins *et al.*, 2015, 2017; Erinosho *et al.*, 2016), the data collected as part of this 3DXRD study can be independently validated. This work aims to establish the 3DXRD method onto the I12 beamline together with the development of a user-friendly data acquisition and processing pipeline.

## 2. Experimental method

### 2.1. Material and microstructure characterization

A single-phase ferritic steel, DX54, with a body-centred cubic (b.c.c.) crystal structure was studied; the composition of the alloy is given in Table 1. This material, supplied by BMW with a thickness of ∼1 mm, is typically used for automotive metal forming applications due to its high ductility. The Zn-galvanized surface, present for environmental protection, was removed using abrasive media. The 3DXRD technique is reliant on individual diffraction spots being distinguished on the detector; this necessitates a limited number of grains diffracting in a single acquisition frame. In conjunction with limits on the incident beam size, determining the probed volume, the grains must be sufficiently coarse to reduce the likelihood of spot overlap on the detector given a relatively large simultaneously illuminated volume of 1 mm × 1 mm (horizontal) × 0.15 mm (vertical). The mean grain size was increased via heat treatment; samples were subjected to an isothermal hold at 980°C for 1 h and cooled at ∼1 K min$^{-1}$. Samples were encapsulated in quartz glass tubes back-filled with Ar. This helped reduce the oxidation during the heat treatment, necessary for the study of a single-phase ferrite.

Table 1
Chemical composition of DX54 steel (Collins *et al.*, 2015).

| Element | Fe | C | P | S | Mn |
|---|---|---|---|---|---|
| wt% | Balance | ≤ 0.06 | ≤ 0.025 | ≤ 0.025 | ≤ 0.35 |

For the characterization of the ferritic steel microstructure, the sample was polished using abrasive media in progressively finer grades, finishing with colloidal silica. The DX54 steel, following heat treatment, was examined using a Zeiss Merlin field emission gun scanning electron microscope (FEG-SEM) equipped with a Bruker e$^-$Flash$^{HR}$ EBSD detector operated with *Esprit 2.0* software. Electron backscatter diffraction (EBSD) data were collected with the microscope operating with a 5 nA probe current and a 20 keV electron beam energy. The maps were taken from an area of 5 mm × 6.5 mm, a region of sufficient size to quantify the texture and grain size of the material with statistical significance.

### 2.2. 3DXRD data collection

Experimental Hutch 1 (EH1) at the I12 beamline of Diamond Light Source (DLS) (Drakopoulos *et al.*, 2015) was used to perform the 3DXRD experiment. A sample with a square rod geometry and a cross section of 1 mm × 1 mm was mounted on the sample stage with the long axis positioned to be axisymmetric with the rotation axis of the sample stage, as shown in Fig. 1(*a*), which describes the intended diffraction geometry.

A multi-distance calibration method (Hart *et al.*, 2013) as implemented in *DAWN* (Basham *et al.*, 2015; Filik *et al.*, 2017) was performed to determine the beam photon energy (60.2 keV) and sample–detector distance, *L* (550.3 mm), using a NIST 674b reference CeO$_2$ calibrant (Cline, 2016) in a dedicated sample holder. Prior to taking all calibration and 3DXRD diffraction measurements, sample alignment with respect to the beam and rotation axis was confirmed using an X-ray imaging camera.

A 1.5 mm (horizontal) × 0.15 mm (vertical) 'letterbox' beam was aligned 0.5 mm above the centre of the sample. For a single letterbox scan, the sample was rotated about the *y* (vertical) axis from −180° to 180° inclusive in steps of 1°. A single 2D diffraction pattern (1 s exposure time) from a Pilatus3 X 2M CdTe detector (pixel size 172 µm) was taken at each step, producing 361 patterns per letterbox scan. Between each letterbox scan, the sample stage was translated by 0.1 mm in *y*, and the sample rotated back to the initial rotation stage angle. A total of ten letterbox scans were collected, enabling all grains to be illuminated within a 1 mm length of the sample.

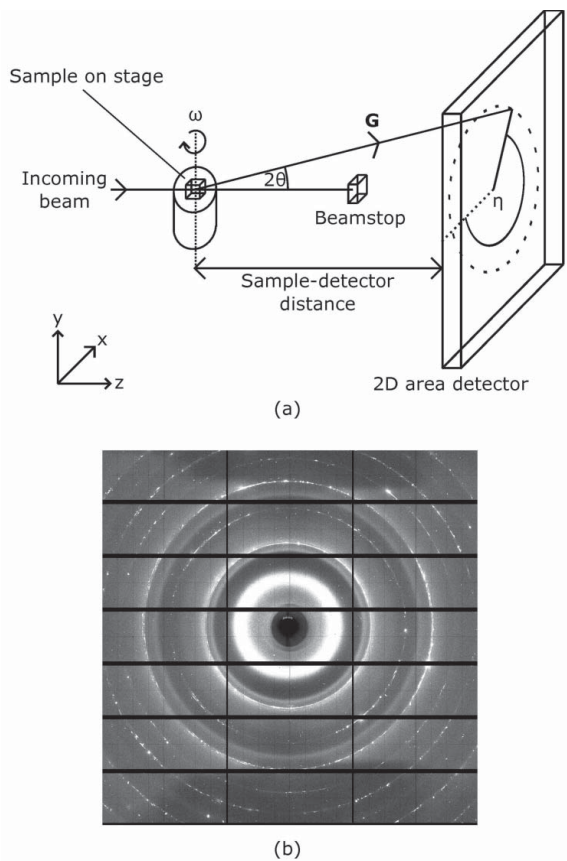

**Figure 1**
3DXRD experimental diagram (*a*) with a typical single-frame diffraction pattern from a 2D area detector (*b*).

### 2.3. 3DXRD analysis

A custom data analysis pipeline for the 2D diffraction patterns was created utilizing algorithms from *ImageD11* for data preparation (Wright, 2020), *GrainSpotter* for initial indexing (Schmidt, 2014) and *FitAllB* for $a_0$ and lattice parameter distortion (Oddershede *et al.*, 2010). All computationally expensive processes were submitted to the Diamond Light Source Scientific Computing cluster (Thorne, 2019) running RedHat Linux 7. Each letterbox scan was indexed in parallel using individual cluster compute nodes, thereby reducing the total indexing time by a factor of ten. The data analysis pipeline is currently run on-demand, after all letterbox scans are collected. The letterbox scan grain maps were then stitched together to generate one complete grain map. Grain analysis scripts were written to automatically generate grain statistics, 3D scatter maps and direct pole figures after indexing.

The data indexing and analysis process is described in detail in Fig. 2, and a full list of all software packages used can be found in the table provided in the supporting information. Peaks that overlap gaps between the detector modules were automatically removed based on whether any pixels belonging to that peak had a brightness value of −1, which would indicate a module gap based on established diffraction image masking criteria at the I12 beamline. Peaks were removed during the cleaning process if they were more than 0.075° away from an expected peak, as calculated by *ImageD11* using the reference unit cell and symmetry.

A number of input parameters are required that determine how *GrainSpotter* assigns individual peaks to grains. For a more complete description of the indexing process, the reader is referred elsewhere (Schmidt, 2014). While the appropriate *GrainSpotter* parameters for each input dataset are well defined mathematically, and may be determined from first principles, in reality the process for determining parameters that yield a satisfying index must be achieved through a highly iterative trial-and-error process. For this dataset, a script was developed that can utilize the high-performance computing cluster at Diamond Light Source to execute multiple *GrainSpotter* runs of the same input dataset in parallel. For each run, the parameters can be either randomly generated, linearly varied within a specified range, or held at a specified value. This significantly increased the rate at which the parameter space was searched. This search was run once on a single letterbox, and desired parameter values were chosen based on the quality of the resultant grain map produced. The grain map quality was judged based on a combination of the number of grains found and the positioning of those grains in 3D space compared with the known illuminated sample volume. The initial *GrainSpotter* index was performed using a published lattice parameter value of $a_0 = 2.8684$ Å (Erinosho *et al.*, 2016). After the initial *GrainSpotter* index, *FitAllB* was used to refine $a_0$ of each letterbox. A geometric filtration on the grains data was performed prior to the refinement to avoid surface effects. After this, individual letterbox $a_0$ values were averaged together to provide a new $a_0$ for the sample. A small change was expected from the published value due to the heat treatment subjected to the ferritic steel studied here. The error on the new $a_0$ was determined by propagating the quoted errors (one standard deviation) on the $a_0$ value calculated by *FitAllB*. The analysis pipeline was repeated with the new $a_0$.

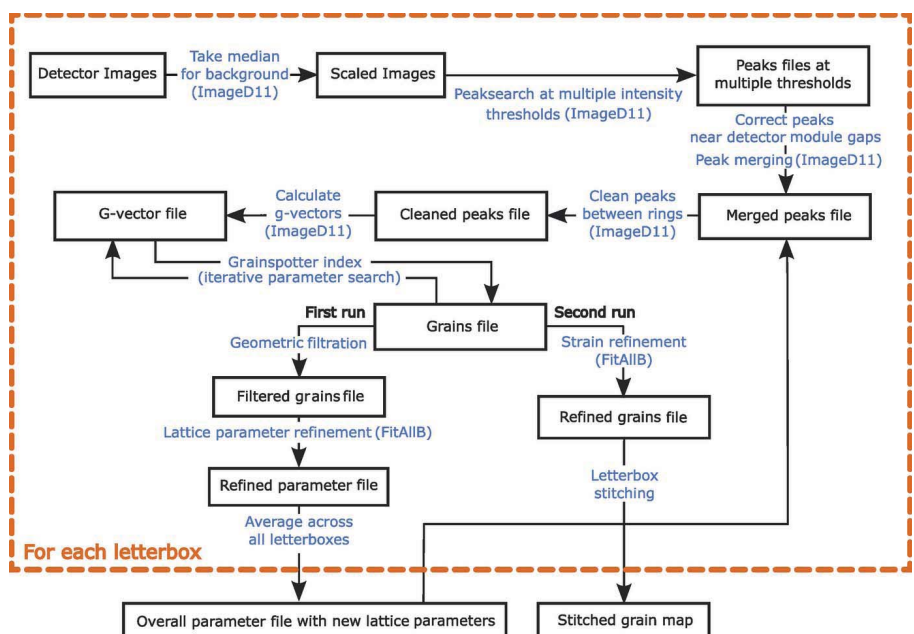

**Figure 2**
3DXRD in-house data processing flowchart.

Each individual letterbox was stitched together using an approach that identifies common grains in the overlap region between the letterboxes. The grain tracking algorithm, described in Algorithm 1, was used with a 50 µm distance tolerance and a 2° orientation tolerance, between grain pairs, to generate a stitched dataset. These parameters determine whether the same grain has been detected in the overlap region between two letterbox scans. First, the grain centre-of-mass positions are offset by the vertical translation of the sample between letterbox scans. Then, a square matrix of distances (`dist_matrix`) between all grain pairs is constructed using the `spatial.distance_matrix` function from the *scipy* Python library (Virtanen *et al.*, 2020). This matrix is masked to True/False values where True indicates a distance that is within the supplied tolerance of 50 µm. A list of grain pairs, `pairs`, is generated by extracting the locations of the True values of the masked `dist_matrix`. Then, for each grain pair in `pairs`, the misorientation between the grains in the pair is calculated using the `disorientation` function from the `crystal.microstructure.Orientation` class of `pymicro` (Proudhon, 2021). This calculation accounts for the cubic symmetries of the grains. If the misorientation is less than the tolerance (2°), that grain pair is added to a list of candidate pairs. Then, for each pair in the list of candidate pairs, the grains in the pair are checked to ensure that they do not originate from the same original letterbox scan. If they pass this check, the smaller of the two grains is selected as the duplicate grain, and is removed from the overall grain database. The grain volume is calculated by *FitAllB* based on the median intensities of the diffracted spots associated with that grain (Oddershede *et al.*, 2010). While this method does not alter the grain data as is output by *FitAllB*, a more accurate technique would involve a volume-weighted average of the different observations of the grain position, orientation and strain. An error in grain position of around 32 µm (half of an average grain radius) is estimated by using this technique over the volume-weighted average technique. Finally, the completed grain database, including offset centre-of-mass positions, can be exported as a contiguous grain map.

---

**Algorithm 1:** Grain tracking algorithm

**input** : database of grain orientations, centre-of-mass positions and originating letterbox scans
**output**: database with duplicated grains removed

1 offset grain positions by sample table vertical stage translation;
2 dist_matrix ← `scipy.spatial.distance_matrix(database)`;
3 dist_matrix_masked ← `MaskLessThanTolerance(dist_matrix, 50 µm)`;
4 pairs ← indices of True values in dist_matrix_masked;
5 **foreach** pair *in* pairs **do**
6    misorien ← `pymicro_disorientation(pair)`;
7    **if** misorien < 2° **then**
8       candidate_pairs.append(pair);
9    **end**
10 **end**
11 **foreach** candidate_pair *in* candidate_pairs **do**
12    **if** *grains in* candidate_pair *do not share origin scans* **then**
13       remove smaller grain of candidate_pair from database;
14    **end**
15 **end**

---

**2.3.1. Error determination.** Errors in grain position, orientation and strain tensor elements were produced by *FitAllB* as a result of the minimization routine. Errors in grain position and strain tensor elements are quoted directly, but orientation errors are further processed. *FitAllB* provides one standard deviation of each component of the Rodrigues vector describing the orientation of the grain relative to the sample. Given a Rodrigues vector of a grain,

$$\mathbf{r} = \tan(\phi/2)\,\mathbf{n} = (r_1, r_2, r_3), \quad (1)$$

and the 'error' vector provided by *FitAllB*,

$$\delta\mathbf{r} = (\delta r_1, \delta r_2, \delta r_3), \quad (2)$$

the error vector is converted into a 3 × 3 rotation matrix, $\mathbf{U}_e$. For small error vectors, this resultant matrix will represent a small rotation close to the 3 × 3 identity matrix. The misorientation angle between $\mathbf{U}_e$ and the identity matrix is calculated using the `Umis` function from the *xfab* Python module (Sørensen *et al.*, 2021). This misorientation angle is treated as the 'error' in the orientation matrix for that grain.

## 3. Results

### 3.1. Microstructure

The grain structure was assessed using EBSD characterization; an inverse pole figure map with respect to the macroscopic direction *z* is shown in Fig. 3. The assessment was made from the same sheet of the annealed DX54 material as the 3DXRD measurements. A mean grain size (spherical equivalent diameter) of 130 µm was found from these data; this was calculated by determining grain boundaries in the microstructure. 1797 grains from low-magnification EBSD scans were used to calculate the mean grain size. Here, the condition for a grain boundary was set as neighbouring pixels with a misorientation of greater than 5°. Grains with fewer than five associated pixels were discarded from the mean grain size calculation. The EBSD data also show that the grain morphology is approximately equiaxed, and, by the preferred blue colouring of the grains in Fig. 3, the material has a strong $\langle 111 \rangle$ texture. This is typical for b.c.c. materials, such as DX54, which are subjected to rolling operations during thermomechanical processing. The distribution of grain orientations is shown more clearly in the corresponding pole figures presented in Fig. 4(*a*).

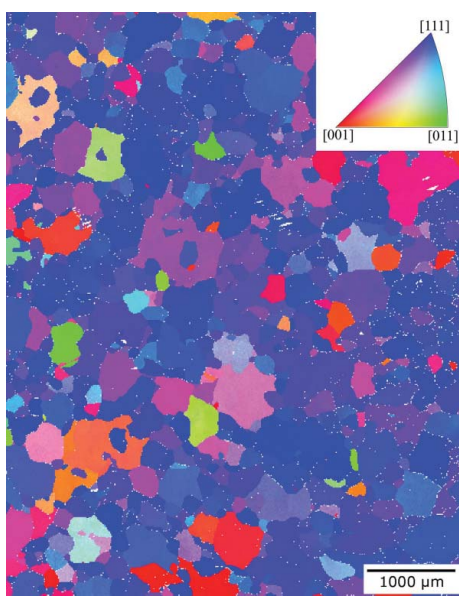

**Figure 3**
Grain orientations measured using EBSD for DX54 ferritic steel, observed on the *xy* (RD-TD) plane of the material. IPF-*z* colouring is shown, where *z* is equivalent to the normal direction, ND.

### 3.2. 3DXRD data

The 3DXRD data were processed using the processing pipeline described earlier, producing for each grain (1) the lattice parameters (corresponding to the basis vectors $|\underline{a}|$, $|\underline{b}|$ and $|\underline{c}|$), (2) strain tensor, (3) orientation, and (4) position. For the 1 mm × 1 mm × 1 mm sample volume illuminated, a total of 1964 grains were identified over the ten letterboxes, with a mean of 196 grains per letterbox. The orientation data for each identified grain were extracted from the 3DXRD dataset, and compared against the EBSD orientation data. This is shown in Fig. 4, generated using the `pymicro` Python library (Proudhon, 2021). Pole figure plots are shown for the $\langle 100 \rangle$, $\langle 110 \rangle$, $\langle 111 \rangle$ and $\langle 310 \rangle$ directions; these are commonly presented for b.c.c. materials to represent texture (*e.g.* Kocks *et al.*, 2000). The average error on a single measurement of the orientation of a grain (one standard deviation) was found to be 0.1°. This relatively small orientation error, coupled with the successful reproduction of the EBSD texture analysis, is suitable evidence that the determination of individual grain orientations with far-field 3DXRD is accurate.

The *GrainSpotter* program is used to estimate the positions of each indexed grain; example results for a single 3DXRD letterbox are shown in Fig. 5. Fig. 5(*a*) demonstrates a typical poor result where the *GrainSpotter* index has been performed with a bad choice of parameters. The output is known to be poor in this example as the positions of many grains lie outside of the volume illuminated by a single letterbox. A drastic difference in the quality of the resultant index can be seen, when compared with the index shown in Fig. 5(*b*). In this case the positions of the grains are found within a realistic cuboidal volume that corresponds well to the illumination volume. Fig. 5(b) represents the result of the iterative *GrainSpotter* parameter search performed for a single letterbox to determine optimized indexing parameters for the rest of the letterbox scans. The average per-grain positional errors (one standard deviation) as reported by *FitAllB* were 101 µm and 36 µm for the horizontal and vertical directions, respectively. In total, 1963 grains were retained in the stitched dataset, representing a total illuminated volume of 1 mm$^3$. Fig. 6 demonstrates the output of the stitching process, with grains coloured by their orientations.

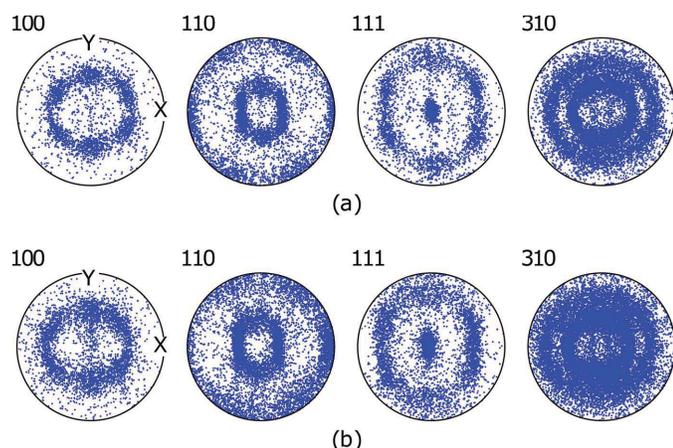

**Figure 4**
Pole figures with orientations plotted from (*a*) EBSD and (*b*) 3DXRD, demonstrating that each method conveys a matching texture.

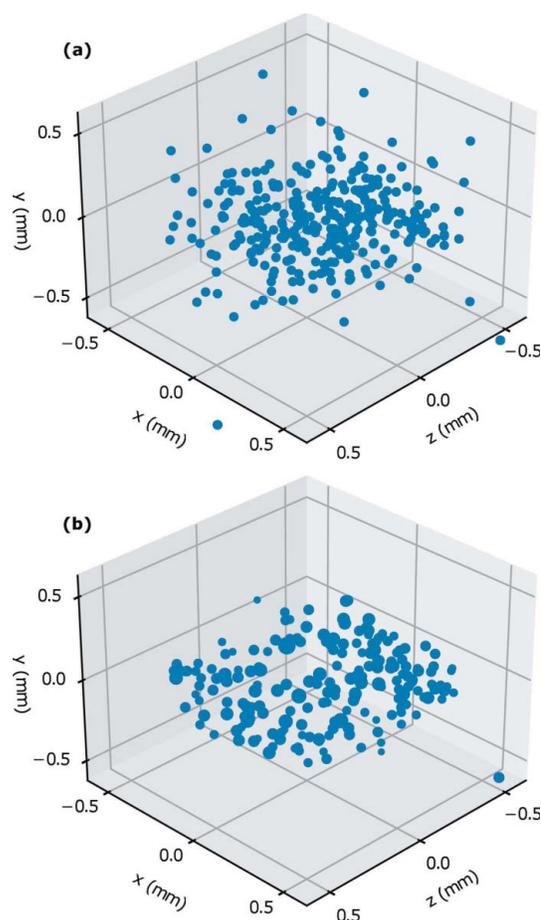

**Figure 5**
Grain centre-of-mass map of one letterbox scan, (*a*) with poor choices for *GrainSpotter* parameters, (*b*) after iterative *GrainSpotter* parameter refinement.

During the refinement process, *FitAllB* provides the lattice distortion tensor for each grain in both the crystal and sample reference frames. This $a_0$ value is firstly refined here, which was determined to be 2.8679(7) Å. This is within the error range for the value previously published for this material (Erinosho *et al.*, 2016), indicating that the heat treatment procedure applied to DX54 in this study had a minimal influence on the lattice parameter. The lattice distortion is presented here as an equivalent elastic hydrostatic strain, $\varepsilon_H$. This is determined by taking the average of the diagonal elements of the lattice distortion tensor. The parameter $\varepsilon_H$ for each grain can be monitored over time as the sample is translated (Juul, 2017). Such inspection is useful as it provides an indication of the stability of the incoming beam energy against time – if the beam energy were to substantially change, the value of $\varepsilon_H$ would be calculated to increase or decrease accordingly due to the observed changes in Bragg diffraction angle. Such an increase or decrease in $\varepsilon_H$ would be incorrect for the DX54 material, as any residual elastic stress distribution would not vary along the long axis of the sample (translation direction). This is performed using grain data from the full 1 mm × 1 mm × 1 mm volume (stitched grain map), shown in Fig. 7, where values of $\varepsilon_H$ are plotted against the grain centre-of-mass height, as per previous studies (Juul, 2017). Here, each data point represents the $\varepsilon_H$ value of a single grain, which is overlaid with a line of best fit. As this fit shows an approximately uniform $\varepsilon_H$ for a given grain centre-of-mass vertical position, there is no beam-related error (such as a shift in beam energy over time) that influences the reliability of grain elastic strain. This is also useful because the beam energy at I12 cannot currently be directly monitored during the course of the scan. A potential solution is to sputter the sample with a powder calibrant such as $CeO_2$ to monitor any beam energy shift.

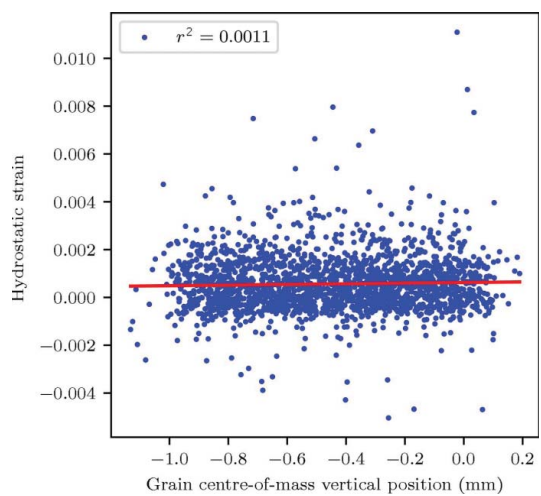

Figure 7
Plot of hydrostatic lattice distortion (hydrostatic strain) variation with grain centre-of-mass vertical position.

The calculated hydrostatic lattice distortion values have also been plotted for every grain for the stitched grain map; this is shown in Fig. 8. The size of each plotted data point is proportional to the grain volume, scaled by the intensity measured for its corresponding diffraction spots.

The degree to which the grain size distribution can be estimated from the 3DXRD dataset is assessed here. For benchmarking, the grain size distribution from the EBSD measurements can be used; the grain sphere equivalent diameter is shown in Fig. 9(a). EBSD grains were smoothed using *MTEX* (Bachmann *et al.*, 2010). Grains with fewer than ten pixels were discarded from the EBSD dataset before plotting. A full description of the EBSD analysis procedure used is available in Appendix A. For the 3DXRD data, the grain volume is proportional to the sum of the intensities of the diffraction peaks associated with that grain (Oddershede *et al.*, 2010). Therefore, taking the cubed root of the intensities sum for each grain creates a distribution that is proportional to the distribution of grain diameters. The resulting 3DXRD distribution is shown in Fig. 9(b).

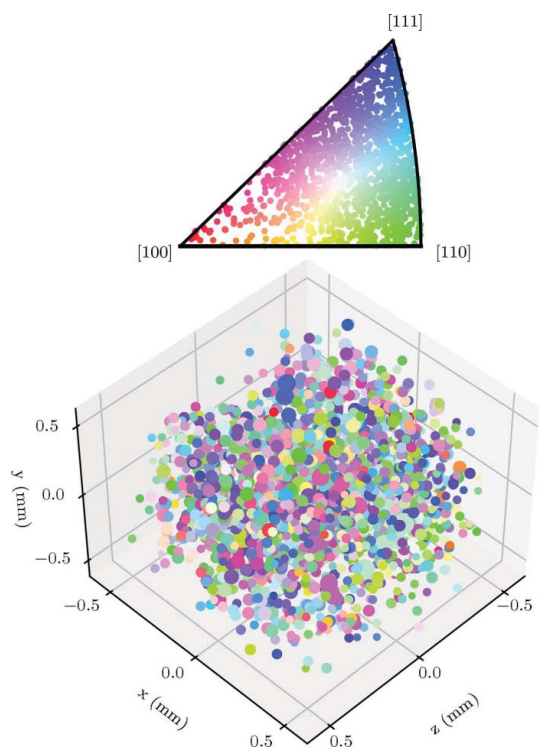

Figure 6
Stitched grain centre-of-mass map of DX54 steel (bottom) with grains coloured by their orientations as per the inverse pole figure (top) relative to the y axis.

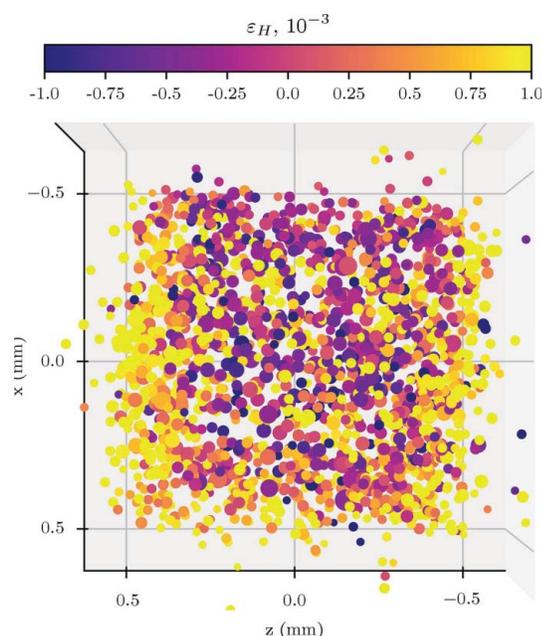

**Figure 8**
Stitched grain centre-of-mass map of DX54 steel, looking down $y$ (top) with grains coloured by the mean of the diagonal elements of the lattice distortion tensor, $\varepsilon_{H}$.

## 4. Discussion

The aim of this study was to determine the feasibility of collecting reliable 3DXRD datasets on the I12 beamline at Diamond Light Source. By collecting data on a microstructurally simple single phase steel, key attributes of the alloy have been obtained, with a sensitivity that is deemed suitable for the study of grain-by-grain dynamic processes. The limitations and constraints of the grain-by-grain orientation measurements, lattice parameters and elastic strain measurements (grain lattice distortion), grain positions and grain size are discussed in turn here. An outlook on the future capability for 3DXRD on the I12 instrument at Diamond is also given.

### 4.1. Grain orientation

The uncertainty of the grain orientation for the Diamond collected data is first considered. An average per-grain orientation error of 0.1° is around twice as large as that found by studies with comparable detector pixel sizes (Bernier *et al.*, 2011; Dake *et al.*, 2016). This is primarily due to the large steps in $\omega$ – steps can be taken to further reduce this error, as discussed in Section 4.5. The successful reproduction of the EBSD measured texture analysis provides good evidence that the determination of individual grain orientations with far-field 3DXRD for data collected on the I12 instrument is accurate. This demonstrates a significant leap forward in non-destructive per-grain texture analysis of relatively thick samples at Diamond Light Source. It shows that an accurate determination of texture from 'spotty' diffraction data from the I12 instrument, with a limited number of discrete grains, is feasible. The primary limiting factors at present for orientation precision are the step size in $\omega$ and the accuracy of the detector calibration.

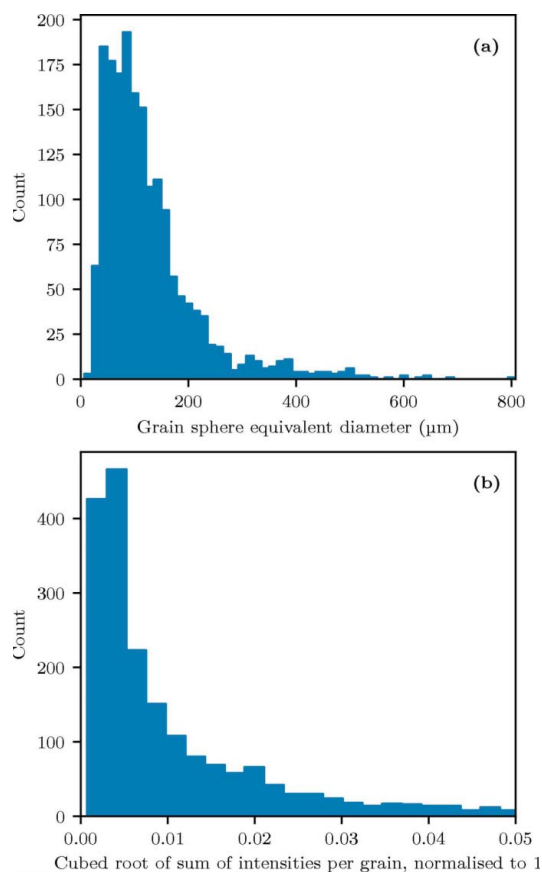

**Figure 9**
Grain size distribution histogram from (*a*) EBSD and (*b*) distribution of the cubed root of grain peak intensities from 3DXRD.

### 4.2. Grain position

While a few outliers can be seen in the grain position map in Fig. 5(*a*), the approximate dimensions of the illuminated volume of one letterbox scan are clearly visible, suggesting a high degree of convergence with *FitAllB*. The grain outliers are primarily caused by *GrainSpotter* mistakenly assigning a collection of peaks to a grain. This could be caused by large tolerance windows in $2\theta$, $\omega$ or $\eta$ space, or a particularly low value for the minimum number of peaks required for a grain to be generated. These parameters were increased to ensure that grains were not overly discarded due to either the increased error in grain position or spots overlapping on the detector. This can be improved by reducing the number of simultaneously illuminated grains (at the cost of data acquisition speed) or by using a detector with a smaller pixel size. Alternatively, a different peak searching procedure, such as a seeded watershed algorithm (Sharma *et al.*, 2012), could be employed that is more adapted to separating peaks that are very close together. This could allow for the detection of a greater number of scattering vectors per grain on average, which would reduce the number of outliers created by *GrainSpotter*. The error in horizontal position provided by *FitAllB* is primarily determined by propagating the experimental errors in peak position and $\omega$ (Oddershede *et al.*, 2010). Therefore a grain position error of around half a detector pixel (pixel size = 172 µm) is expected here, and comparable with other 3DXRD studies (which typically use detectors with much smaller pixel sizes) (Poulsen, 2012; Nervo *et al.*, 2014; Renversade & Borbély, 2017). A significant difference between the horizontal and vertical positional errors was observed, as in other 3DXRD experiments, such as that by Nervo *et al.* (2014). In that paper, the discrepancy was attributed to differences in systematic error between the horizontal and vertical axes, as the vertical position of each grain is roughly constant during rotation.

The problems caused by detector spot overlap can be reduced by defining a smaller X-ray beam height, and stitching together multiple grain maps (collected at different sample vertical translations) during the data analysis process. This procedure increases the data acquisition time, but allows for a large number of grains to be indexed overall, to improve bulk statistics for analyses of properties such as sample texture. For this experiment, ten grain maps were collected and later stitched together to form an overall grain map, as per Fig. 8. While a few outliers can be seen, the approximate dimensions of the sample are clearly visible in the grain map, and the outliers could be easily removed by a bounding box filtration routine. It is clear from the stitched grain map how 3DXRD as a technique gives unparalleled access to bulk grain positional information for *in situ* experiments. This shows promise for obtaining rich datasets in further 3DXRD experiments at Diamond Light Source.

### 4.3. Grain size

The grain size distribution is approximately log-normal in both the EBSD and 3DXRD datasets (see Fig. 9) which is expected for a single-phase ferritic alloy like DX54. A lack of smaller grains is evident in the EBSD distribution compared with the 3DXRD distribution. This is likely due to the removal of small grains from the EBSD dataset prior to the grain size distribution analysis.

### 4.4. Grain lattice distortion

The average grain lattice distortion tensor element error in the two horizontal directions ($xx$ and $zz$) in Table 2 are approximately equal, for the same reasons as the similar effect observed in grain positional error. The errors in relative lattice distortion are quite large compared with the recorded lattice distortion values, and errors achieved by other 3DXRD beamlines, such as $1 \times 10^{-5}$ at beamline ID11 of the ESRF (Oddershede *et al.*, 2010). The sources of these errors are described in more detail in Section 4.5. Because of these factors, individual grain lattice parameter measurements are not generally useful due to the large relative error, but general trends can be inferred. An example of where this has been successful in this dataset is the identification of different grain lattice distortions between grains towards the $z$-axis edges and grains in the bulk in Fig. 8, which is explained in Section 4.6.

**Table 2**
Sample-average errors in grain strain tensor diagonal elements as output by *FitAllB*.

| Strain tensor element | Sample-average error in element ($\times 10^{-3}$) |
|---|---|
| $\varepsilon_{xx}$ | 1.5 |
| $\varepsilon_{yy}$ | 0.6 |
| $\varepsilon_{zz}$ | 1.6 |

### 4.5. Resolution limitations

There are a number of current limitations that affect the precision of the grain-resolved position, orientation and strain:
(1) Detector pixel size and point spread function.
(2) Sub-optimal calibration of experimental geometry.
(3) Large step size in $\omega$ (1°) – according to Oddershede *et al.* (2010), if a peak only appears in one or two images due to the large step size, the assumption that errors are Gaussian cannot be entirely fulfilled.
(4) Uncorrected module misalignments in the detector.
(5) Higher than expected powder background affecting peak searching efforts.

There are solutions to some of these factors, which are in the process of being implemented at I12:
(1) Experimental geometry can be more accurately calibrated with the use of a single-crystal diffraction standard (Wong-Ng *et al.*, 2001).
(2) The step size in $\omega$ can be reduced for subsequent experiments via the implementation of hardware-triggered fly scanning of the rotation table (now implemented).
(3) Characterizing the detector module misalignments (and correcting the peaks accordingly) via the acquisition of multiple $CeO_2$ patterns with different beam centres (Wright *et al.*, 2022).

With these changes implemented, grain scattering vectors could be more accurately determined, which would significantly improve the resolution in strain, and somewhat improve the resolution in orientation and position.

### 4.6. Raw peaks

The first frame from the first letterbox scan is shown in Fig. 1(*b*). While some spot overlap is visible, a large number of discrete spots are observed, and the individual diffraction rings are clearly seen. This demonstrates good choices for both the illuminated sample volume and sample–detector distance. The broad bright halo towards the centre of the pattern was caused by low-angle X-ray scattering through the amorphous carbon window on the sample holder. This low-angle scattering was removed during the indexing procedure so can be safely ignored.

An interesting feature of the alloy studied was the presence of satellite peaks present at the leading and trailing edges of the ferrite reflections, as seen in the example {110} reflection, Fig. 10(*a*). The exact source of these peaks is currently unknown, but it is suspected that X-ray scattering upstream of the sample could be the cause. This is currently under further investigation. Although the unexpected peaks were mostly filtered by peak-searching at multiple intensity thresholds, some peaks may have remained that could be unintentionally indexed during the data analysis procedure. Fig. 10(*b*) demonstrates the grain peaks for the same letterbox scan post-filtration.

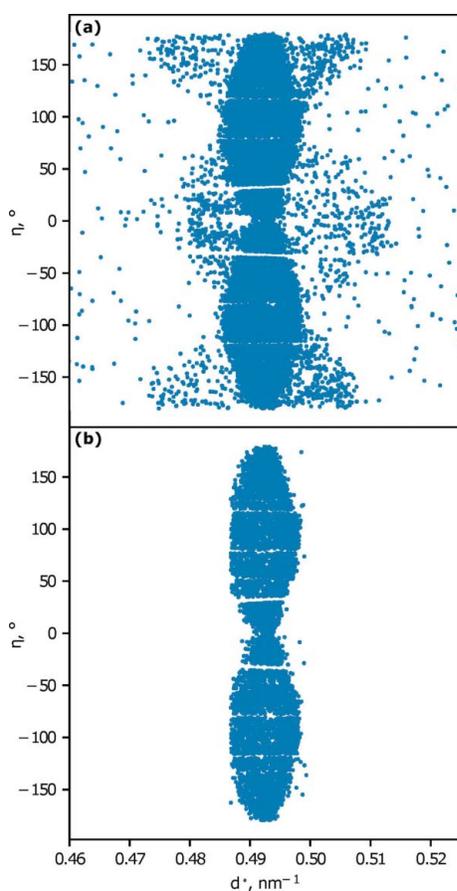

**Figure 10**
Radially integrated {110} plot of peaks from one letterbox scan (*a*) before filtration and (*b*) after filtration.

The $\varepsilon_H$ plots in Fig. 8 show some grains on the $z = \pm 0.5$ surfaces appearing with an apparently high lattice distortion ($1 \times 10^{-3}$) relative to the refined $a_0$ value determined by *FitAllB*. These originally galvanized surfaces (the *x*–*y* surfaces in the sample grain plots) were mostly removed during sample preparation; however, a small amount of Zn remained which diffused into the surface during the annealing heat treatments. The grains in the vicinity have a modified lattice parameter compared with the bulk alloy due to the Zn impurity (Marder, 2000). This is shown most clearly by investigating the lattice parameters calculated from the diagonal lattice distortion tensor elements. Fig. 11 shows that the lattice parameters at the sample edges are slightly different than the lattice parameter in the bulk of the material. No peak broadening or smearing was observed from these surface grains.

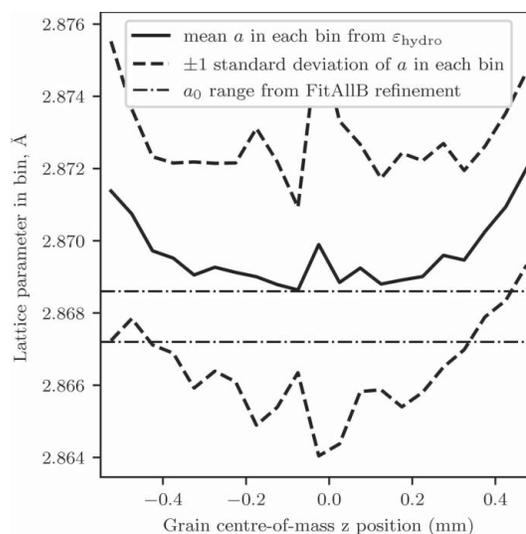

**Figure 11**
Plot of lattice parameter variation calculated from the grain hydrostatic lattice distortion (binned in *z* with a bin width of 0.05 mm) with grain centre-of-mass position along *z*. Lattice parameter error bars are standard deviations within the bins. The $a_0$ range represents the value obtained from the *FitAllB* $a_0$ refinement ± one standard deviation.

### 4.7. Outlook

The ability to successfully index a significant number of grains in this experiment establishes a wide range of possibilities for future 3DXRD experiments at the I12 beamline, such as *in situ* deformation, thermal treatment or fatigue studies, which are all regularly carried out at the beamline. The flexibility and adaptability of the equipment and experimental layout of the I12 beamline lends itself very well to *in situ* measurements that have previously been impossible to perform on a per-grain level. The combination of *in situ* 3DXRD experiments at I12 and *ex situ* 'post-mortem' sample studies, utilizing techniques such as diffraction-contrast tomography or scanning 3DXRD at other beamlines, creates a powerful multi-facility data collection routine for the analysis of complicated materials. Further developments of the DLS 3DXRD data analysis pipeline are ongoing, with explorations into the indexing of multi-phase materials, the adoption of new peak searching routines, and the analysis of *in situ* deformation studies.

### 5. Conclusion

A study was conducted to determine the feasibility for 3DXRD experiments at the I12 beamline at Diamond Light Source, demonstrating this on a polycrystalline low-carbon ferritic steel.

A demonstration far-field 3DXRD experiment on the microstructually simple ferritic steel DX54 was performed on the I12 beamline; around 2000 grains within a 1 mm³ region were successfully identified. The position, orientation, strain and relative volume of each of these grains were determined with indexing and refinement software. This indicates per-grain characteristics necessary for detailed grain-by-grain studies of phenomena in polycrystals is feasible at this instrument.

An analysis pipeline created for I12-generated 3DXRD data has been developed, which integrates established computational tools and software that are widely used by the existing 3DXRD user community. Additionally, a grain stitching procedure was created which combines data from scans that have a constrained probed volume subset, to study larger polycrystal agglomerates.

Grain orientations were determined to an average error of $\sim 0.1°$ (one standard deviation). This allowed the generation of pole figures that are in good agreement with EBSD data collected on the same alloy – this is a significant result for this experiment and shows that high-quality texture measurements on 'spotty' X-ray diffraction data is possible using our 3DXRD analysis pipeline.

Grain centre-of-mass positions were determined to an accuracy of $\sim 80 \, \mu m$ (horizontally, one standard deviation). The sensitivity of initial indexing quality to *GrainSpotter* parameter changes was demonstrated – future experiments should explore methods to optimize this processing step. Inferring the distribution of grain masses inferred from the grain-intensity distribution also showed promising results.

Residual elastic grain strains were obtained with an error of $\sim 1 \times 10^{-3}$. This can be reduced during future experiments by reducing the $\omega$ angular step size, which is recommended for *in situ* 3DXRD experiments studied on the I12 beamline.

The grain size distribution from the EBSD dataset was shown to be as expected for this alloy, and in good agreement with the shape of the 3DXRD grain peak intensity distribution, after accounting for small grains discarded during the EBSD analysis.

## 6. Related literature

The following references, not cited in the main body of the paper, have been cited in the supporting information: Collette *et al.* (2021); Harris *et al.* (2020); Hunter (2007); de Jager (2021); Knudsen *et al.* (2013); McKinney (2010); Van Kemenade *et al.* (2021).

## APPENDIX *A*
### EBSD grain size analysis procedure

The procedure used to extract grain sphere equivalent diameters from EBSD data is as follows:

(1) Import EBSD data into *MTEX*.
(2) Segment EBSD data into grains based on a 5° misorientation tolerance.
(3) Remove grains with fewer than ten contributory pixels.
(4) Re-segment EBSD data with the same 5° misorientation tolerance.
(5) Interpolate grain boundary coordinates with *MTEX* `grains.smooth` function.
(6) Generate array of grain sphere equivalent radii from *MTEX* `grains.equivalentRadius` function.

(7) Double each element in the array to generate an array of grain sphere equivalent diameters.

### Acknowledgements

This work was carried out with the support of Diamond Light Source, Instrument Beamline I12 (proposal NT26376). The authors are grateful to Bonnie Attard, Robert Atwood, Simon Cannon, Sam Cruchley, Bob Humphreys, Jette Oddershede, Mary Taylor and Jon Wright for their assistance and advice on sample preparation, data analysis and preparation of this manuscript.

### Funding information

Funding for this research was provided by Diamond Light Source and the University of Birmingham (studentship to James A. D. Ball). Anna Kareer acknowledges the support from the Engineering and Physical Sciences Research Council under Fellowship grant EP/R030537/1.

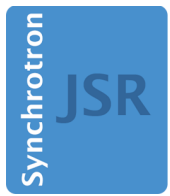

**Volume 29 (2022)**

**Supporting information for article:**

Implementing and evaluating far-field 3D X-ray diffraction at the I12 JEEP beamline, Diamond Light Source

**James A. D. Ball, Anna Kareer, Oxana V. Magdysyuk, Stefan Michalik, Anastasia Vrettou, Neal Parkes, Thomas Connolley and David M. Collins**

# Implementing and Evaluating Far-Field 3D X-Ray Diffraction at the I12 JEEP Beamline, Diamond Light Source

James A. D. Ball[a,b], Anna Kareer[c], Oxana V. Magdysyuk[b], Stefan Michalik[b], Anastasia Vrettou[a], Neal Parkes[a], Thomas Connolley[b], and David M. Collins [*a]

[a]School of Metallurgy and Materials, University of Birmingham, Edgbaston, Birmingham B15 2TT, United Kingdom
[b]Diamond Light Source Ltd., Harwell Science and Innovation Campus, Didcot OX11 0DE, United Kingdom
[c]Department of Materials, University of Oxford, Oxford, OX1 3PH, United Kingdom

# Supplementary Content

The table below lists each software package that was used in the data analysis of 3DXRD data collected from the I12 beamline, Diamond Light Source.

Table 1: Software package usage in data analysis process.

| Software package | Reference | Usage |
| --- | --- | --- |
| ImageD11 | Wright (2020) | Peak searching, merging, cleaning |
| xfab | Sørensen *et al.* (2021) | Orientation error determination |
| GrainSpotter | Schmidt (2014) | Initial index |
| FitAllB | Oddershede *et al.* (2010) | Refinement of lattice parameter and strain |
| numpy | Harris *et al.* (2020) | All stages |
| pandas | McKinney (2010) | Internal database management |
| matplotlib | Hunter (2007) | Graph plotting |
| pymicro | Proudhon (2021) | Grain tracking |
| scipy | Virtanen *et al.* (2020) | Letterbox stitching |
| h5py | Collette *et al.* (2021) | Detector image processing |
| Pillow | Van Kemenade *et al.* (2021) | Detector image processing |
| fabio | Knudsen *et al.* (2013) | Detector image processing |
| jsmin | de Jager (2021) | Processing input files |
| MTEX | Bachmann *et al.* (2010) | Grain volume distribution |

*d.m.collins@bham.ac.uk